# Plasmonic generation of spin currents

K. Uchida[1,2*], H. Adachi[3,4], D. Kikuchi[1,5], S. Ito[1], Z. Qiu[5], S. Maekawa[3,4] and E. Saitoh[1,3-5]

[1]*Institute for Materials Research, Tohoku University, Sendai 980-8577, Japan,*
[2]*PRESTO, Japan Science and Technology Agency, Saitama 332-0012, Japan,*
[3]*Advanced Science Research Centre, Japan Atomic Energy Agency, Tokai 319-1195, Japan,*
[4]*CREST, Japan Science and Technology Agency, Tokyo 102-0075, Japan,*
[5]*WPI Advanced Institute for Materials Research, Tohoku University, Sendai 980-8577, Japan.*
[*]e-mail: kuchida@imr.tohoku.ac.jp

**Surface plasmons, free-electron collective oscillations in metallic nanostructures, provide abundant routes to manipulate light-electron interactions that can localize light energy and alter electromagnetic field distributions at subwavelength scales. The research field of plasmonics[1-5] thus integrates nano-photonics with electronics. In contrast, electronics is also entering a new era of spintronics[6,7], where spin currents[8,9] play a central role in driving devices. However, plasmonics and spin-current physics have so far been developed independently. Here, we demonstrate the generation of spin currents by surface plasmons. Using Au nanoparticles embedded in a $Pt/BiY_2Fe_5O_{12}$ bilayer film, we show that, when the Au nanoparticles fulfill the surface-plasmon-resonance condition, spin currents are generated across the $Pt/BiY_2Fe_5O_{12}$ interface. This plasmonic spin pumping results from nonequilibrium states of spins excited by surface-plasmon-induced evanescent electromagnetic fields in the $BiY_2Fe_5O_{12}$ layer. Such plasmonic spin pumping will invigorate research on spin-current physics, and pave the way for future spin-based plasmonic devices.**

In the field of spintronics, for spin-current generation, we often rely on a spin pumping effect[10-13], which refers to the transfer of spin-angular momentum from magnetization dynamics in a ferromagnet to conduction-electron spins in an attached paramagnet; when a motion of magnetic moments in the ferromagnet is excited, a spin current is pumped out of the ferromagnet into the paramagnet.

Recent studies in spintronics have revealed that there are two types of spin pumping effects. One is coherent spin pumping, which is due to coherent phase precession of the magnetization, i.e., ferromagnetic or spin-wave resonances in ferromagnets typically excited by microwave irradiation[11-13]. The other is incoherent spin pumping, which is due to a nonequilibrium state of spins at a ferromagnet/paramagnet interface excited by incoherent external perturbations; when magnons in the ferromagnet and/or electrons in the paramagnet deviate from thermal equilibrium, a net spin current is generated across the interface. A typical example of the incoherent spin pumping is the spin Seebeck effect[14-21], where the magnon/electron distribution functions are modulated by external temperature gradients.

In this work, we report the observation of *plasmonic* spin pumping; we show that the incoherent spin pumping can be driven by surface plasmons under visible light illumination. This plasmonic spin pumping enables the conversion of light into a pure spin current, thereby paving the way to create new devices, such as solar spin-current generators.

Figure 1a shows a schematic illustration of the sample system used in the present study. The sample consists of a paramagnetic Pt/ferrimagnetic $BiY_2Fe_5O_{12}$ bilayer film with Au nanoparticles (NPs) embedded in the $BiY_2Fe_5O_{12}$ layer. The Au NPs with the in-plane diameter of 30-90 nm and the height of 20-50 nm were formed by heating a Au thin film[22] on a single-crystalline $Gd_3Ga_5O_{12}$ (111) substrate (Fig. 1c). The 110-nm-thick $BiY_2Fe_5O_{12}$ film was then coated on them by means of a metal-organic decomposition (MOD) method[20] (see Methods). Finally, a 5-nm-thick Pt film was sputtered on the $BiY_2Fe_5O_{12}$ film. To excite surface plasmons, the Pt/$BiY_2Fe_5O_{12}$/Au-NP sample was illuminated with unpolarized monochromatic light with the wavelength $\lambda$ (400-770 nm) and power $P$ from the Pt-film side at normal incidence (Fig. 1a). Since the Pt and $BiY_2Fe_5O_{12}$ layers are very thin, the light passes through the layers and interacts with Au NPs. We checked that the electric resistance between the ends of the $BiY_2Fe_5O_{12}$ layer with Au NPs under the light illumination is much greater than the measurement limit of our electrometer ($>2 \times 10^{11}$ Ω) at all the $\lambda$ values; the possibility of optical charge-carrier generation in $BiY_2Fe_5O_{12}$ is excluded.

In Fig. 2a, we show the light-transmittance spectrum of the Pt/$BiY_2Fe_5O_{12}$/Au-NP sample. We found that the sample exhibits clear dip structure in the transmittance spectrum around $\lambda = 690$ nm in comparison with the spectrum of a Pt/$BiY_2Fe_5O_{12}$ sample without Au NPs. This dip structure is attributed to the excitation of localized surface plasmon resonance (SPR)[1,2] in Au NPs embedded in the $BiY_2Fe_5O_{12}$ layer. To visualize the SPR in the $BiY_2Fe_5O_{12}$/Au-NP structure, we also performed electromagnetic field simulations by means of a finite-difference time-domain (FDTD) method[23]; as shown in Fig. 1d, under the SPR condition ($\lambda = 690$ nm), strong evanescent electromagnetic fields (near fields)[1,2] are generated in the $BiY_2Fe_5O_{12}$ film in the vicinity of the $BiY_2Fe_5O_{12}$/Au-NP interface (see Supplementary Information (SI) section A1).

The measurement mechanism of the plasmonic spin pumping is as follows. If the evanescent electromagnetic fields concomitant with surface plasmons in Au NPs excite magnons in the $BiY_2Fe_5O_{12}$ film and the excited magnons drive the incoherent spin pumping, a spin current is generated in the Pt layer with the spatial direction $\mathbf{J}_S$ and the spin-polarization vector $\boldsymbol{\sigma}$ parallel to the magnetization $\mathbf{M}$ of the $BiY_2Fe_5O_{12}$ film (Fig. 1b). This plasmon-induced spin current is converted into a d.c. charge current $\mathbf{J}_C$ due to the inverse spin Hall effect (ISHE)[12,24,25] in the Pt layer owing to the strong spin-orbit interaction in Pt. When $\mathbf{M}$ of the $BiY_2Fe_5O_{12}$ film is along the $x$ direction, $\mathbf{J}_C$ is generated along the $y$ direction because of the following ISHE symmetry:

$$\mathbf{J}_C \propto \mathbf{J}_S \times \boldsymbol{\sigma} \tag{1}$$

where the $x$, $y$, and $z$ directions are defined in Fig. 1b. We note that extrinsic artifacts induced by a static magnetic proximity effect[26] at the Pt/$BiY_2Fe_5O_{12}$ interface are negligibly small in the present structure (see ref. 21 and SI section B). To detect the ISHE induced by the plasmonic spin pumping, we measured an electric voltage $V$ between the ends of the Pt layer while illuminating with monochromatic light and applying an external magnetic field $\mathbf{H}$ (with magnitude $H$) to the Pt/$BiY_2Fe_5O_{12}$/Au-NP sample along the $x$-$y$ plane (Fig. 1a). All of the $V$ measurements were carried out at room temperature and atmospheric pressure.

To observe the plasmonic spin pumping, it is important to separate the plasmon-induced signals from extrinsic heating effects. First of all, to check the heating effect due to the light illumination, we measured the voltage in the Pt/$BiY_2Fe_5O_{12}$ sample without Au NPs. Figure 2b shows the $\lambda$ dependence of the voltage normalized by the incident light power, $V/P$, in the Pt/$BiY_2Fe_5O_{12}$ sample at $H = 200$ Oe. When $\mathbf{H}$ is applied along the $x$ direction, a finite voltage signal appears in the Pt layer at all the $\lambda$ values (see gray triangle data points in Fig. 2b). Because of the absence of Au NPs and surface plasmons, the signal in the Pt/BiY$_2$Fe$_5$O$_{12}$ sample is attributed to the heating of the sample by the light illumination, i.e., the longitudinal spin Seebeck effect[18,20,21] (see also SI section C). Here, the sign of the $V/P$ signal shows that the temperature of the Pt layer is higher than that of the BiY$_2$Fe$_5$O$_{12}$ layer under the light illumination[18]. Notably, this heating signal in the Pt/BiY$_2$Fe$_5$O$_{12}$ sample does not exhibit strong $\lambda$ dependence (Fig. 2b).

Now we are in a position to investigate the plasmonic generation of spin currents. The blue circle data points in Fig. 2b are the $V/P$ spectrum in the Pt/BiY$_2$Fe$_5$O$_{12}$ film that contains Au NPs. Also in this Pt/BiY$_2$Fe$_5$O$_{12}$/Au-NP sample, when $\mathbf{H} \parallel x$, the $V/P$ signal appears with the same sign (Fig. 2b) and the magnitude of $V$ is proportional to the light power $P$ (Fig. 2c). The sign of the signal at each $\lambda$ was observed to be reversed by reversing $H$ with a hysteresis loop, indicating that the signal is affected by the $\mathbf{M}$ direction of the BiY$_2$Fe$_5$O$_{12}$ layer (Fig. 2d). As also shown in Fig. 2b, we confirmed that the $V/P$ signal disappears when $\mathbf{H}$ is applied along the $y$ direction, consistent with equation (1). These results indicate that the observed $V/P$ signal in the Pt/BiY$_2$Fe$_5$O$_{12}$/Au-NP sample is due to the ISHE induced by spin currents in the Pt layer. Significantly, Fig. 2b shows that the $V/P$ signal in the Pt/BiY$_2$Fe$_5$O$_{12}$/Au-NP sample exhibits peak structure and is dramatically enhanced around $\lambda = 690$ nm. This peak position in the $V/P$ spectrum coincides with the SPR wavelength of this Pt/BiY$_2$Fe$_5$O$_{12}$/Au-NP sample (compare Figs. 2a and 2b). Since both the Pt/BiY$_2$Fe$_5$O$_{12}$/Au-NP and Pt/BiY$_2$Fe$_5$O$_{12}$ samples are illuminated from the Pt-film side, the temperature rise of the Pt layer is almost the same for both, confirming that the heating of the Pt layer is irrelevant to the ISHE enhancement under the SPR condition in the Pt/BiY$_2$Fe$_5$O$_{12}$/Au-NP sample (note again that the background heating signal does not exhibit strong $\lambda$ dependence). This peak $V/P$ structure in the Pt/BiY$_2$Fe$_5$O$_{12}$/Au-NP sample also cannot be explained by the heating of Au NPs or the BiY$_2$Fe$_5$O$_{12}$ layer under the SPR condition because of the different sign; if the BiY$_2$Fe$_5$O$_{12}$ layer was heated by electromagnetic loss in Au NPs, the ISHE voltage should decrease under the SPR condition since the temperature gradient generated by the Au-NP heating across the Pt/BiY$_2$Fe$_5$O$_{12}$ interface is of an opposite sign to the case when the Pt layer is heated (see also Fig. 3). Possible small heating of Au NPs in the off-resonance region might reduce the magnitude of the background signal slightly, but is also irrelevant to the $V/P$ signal with the positive peak structure. Since similar $V/P$ signals were observed in a Au/BiY$_2$Fe$_5$O$_{12}$/Au-NP sample, where the Pt film is replaced with a 5-nm-thick Au film, proximity ferromagnetism in Pt also cannot be the origin of this peak voltage structure (see SI section B). In contrast, this voltage signal was found to disappear in a Cu/BiY$_2$Fe$_5$O$_{12}$/Au-NP sample, where the Pt layer is replaced with a 5-nm-thick Cu film in which the spin-orbit interaction is very weak, indicating the important role of spin-orbit interaction, or the ISHE, in the voltage generation (Fig. 2e). The voltage signal also disappears in a Pt/Gd$_3$Ga$_5$O$_{12}$/Au-NP sample, where Au NPs are embedded in a paramagnetic insulator Gd$_3$Ga$_5$O$_{12}$ film instead of BiY$_2$Fe$_5$O$_{12}$ (see Methods), indicating that direct contact between BiY$_2$Fe$_5$O$_{12}$ and Pt is necessary for the observed voltage generation; electromagnetic artifacts are irrelevant (Fig. 2f). Our numerical calculations of electromagnetic field distributions based on the FDTD method show that the electromagnetic coupling between the Pt film and Au NPs under the SPR condition is too weak to explain the enhancement of the ISHE signal in the Pt/BiY$_2$Fe$_5$O$_{12}$/Au-NP sample (see SI section A2).

To further verify the effect of surface plasmons on the spin-current generation, we measured the $V/P$ spectra in the Pt/BiY$_2$Fe$_5$O$_{12}$/Au-NP samples with changing the contact condition between BiY$_2$Fe$_5$O$_{12}$ and Au

NPs. Here, we prepared two Pt/BiY$_2$Fe$_5$O$_{12}$/Au-NP samples by different processes A and B (samples A and B, detailed in Methods). In the Pt/BiY$_2$Fe$_5$O$_{12}$/Au-NP sample A, which was used for the experiments in Fig. 2, Au NPs are densely embedded in the BiY$_2$Fe$_5$O$_{12}$ film (Fig. 3e), where the peak *V/P* structure appears under the SPR condition (Figs. 3b and 3d). In contrast, in the sample B, Au NPs are separated from the BiY$_2$Fe$_5$O$_{12}$ film by small voids (Fig. 3j), which interrupt the interaction between magnons in BiY$_2$Fe$_5$O$_{12}$ and surface plasmons in Au NPs since the evanescent electromagnetic fields generated by the SPR are localized in the vicinity of Au NPs (see SI section A3). In the Pt/BiY$_2$Fe$_5$O$_{12}$/Au-NP sample B, dip structure was found to appear in the *V/P* spectrum under the SPR condition (Figs. 3g and 3i). This behaviour was observed not only in one sample but also in our different samples as exemplified in the insets to Figs. 3d and 3i. The dip *V/P* signal in the Pt/BiY$_2$Fe$_5$O$_{12}$/Au-NP sample B is attributed to the heating of Au NPs by the SPR, which reduces the temperature gradient across the Pt/BiY$_2$Fe$_5$O$_{12}$ interface and the longitudinal spin Seebeck voltage (see also SI section A3). These results indicate that the enhancement of the ISHE under the SPR condition requires direct contact between BiY$_2$Fe$_5$O$_{12}$ and Au NPs. This is evidence for the operation of the plasmonic spin pumping due to nonequilibrium magnons excited by surface plasmons.

The above experiments clearly show that the spin current generated by the light illumination in the Pt/BiY$_2$Fe$_5$O$_{12}$/Au-NP sample A is strongly enhanced under the SPR condition and this enhancement is not connected to extrinsic heating effects. So then, what is the origin of the spin current under the SPR condition? A strong candidate for the mechanism that enhances the ISHE voltage is the excitation of magnons by the strong evanescent electromagnetic fields (near-field photons) concomitant with surface plasmons. If the near-field photons excite magnons via the magnon-photon interaction[27-30] near the BiY$_2$Fe$_5$O$_{12}$/Au-NP interface, the excited magnons in the BiY$_2$Fe$_5$O$_{12}$ layer produce a spin current owing to the incoherent spin-pumping mechanism and the positive ISHE voltage in the Pt layer: this is the plasmonic spin pumping. We formulate theoretically the plasmonic spin pumping using a standard many-body technique. The spin current pumped by light illumination is described as

$$J_\mathrm{S} = \Gamma^+ \Gamma^- G_\mathrm{S} \frac{\hbar \nu a_\mathrm{S}^4}{c\alpha} |E|^2 \qquad (2)$$

within the linear response region (see SI section D for details). Here, $\Gamma^{+(-)}$ is the dimensionless magnon-photon coupling constant, $G_\mathrm{S}$ the strength of the magnetic coupling at the Pt/BiY$_2$Fe$_5$O$_{12}$ interface, $\hbar$ the Planck constant divided by $2\pi$, $\nu$ the photon frequency, $c$ the velocity of light, $a_\mathrm{S}^3$ the effective block spin volume, $\alpha$ the Gilbert damping constant, and $E$ the electric field amplitude induced by surface plasmons. Equation (2) indicates that the plasmon-induced spin current in the Pt/BiY$_2$Fe$_5$O$_{12}$/Au-NP sample is proportional to the electromagnetic energy in the BiY$_2$Fe$_5$O$_{12}$ layer ($\propto |E|^2$), a situation consistent with the experimental results in Fig. 2c since $|E|^2$ generated by surface plasmons is proportional to the incident light power *P*. The observed peak ISHE voltage in the Pt/BiY$_2$Fe$_5$O$_{12}$/Au-NP sample A is attributed to the strong evanescent electromagnetic fields enhanced by surface plasmons, while the spin current described by equation (2) is undetectably small in the plain Pt/BiY$_2$Fe$_5$O$_{12}$ film due to the small energy transfer from external photons to magnons (see SI section D and note that the ratio of the voltage enhancement in Fig. 2b does not correspond to that of the $|E|^2$ enhancement due to the presence of the background heating signal).

The plasmonic spin pumping observed here can be applied to the construction of light/spin-current convertors for driving spintronic devices, thereby opening the door to "plasmonic spintronics". The plasmonic spin pumping also adds a light/voltage conversion function to spin Seebeck thermoelectric devices[20], enabling hybrid voltage generation from both light and heat in a single device. The plasmonic spin pumping follows the scaling law same as the spin Seebeck effect: the output power is increased simply by extending the device area. This light/voltage conversion is conceptually different from that based on a PN junction, and is free from the output-voltage limitation caused by built-in potential. Although the spin and electric voltages generated by the plasmonic spin pumping are still small, there is plenty of scope for performance improvement; they can be enhanced by (1) using magnetic insulators with large magnon-photon coupling, (2) improving magnetic-insulator/NP interfaces for efficient magnon-plasmon energy transfer, and (3) introducing plasmonic crystals that exhibit sharp and strong SPR.

**Methods**

**Preparation process of Pt/BiY$_2$Fe$_5$O$_{12}$/Au-NP samples.** The Au NPs were formed by the following two-step procedures: (1) fabrication of a 5-nm-thick Au thin film on a Gd$_3$Ga$_5$O$_{12}$ (111) substrate by using a d.c. sputtering system and (2) heating of the Au film at 1000 °C for 30 minutes[22]. To increase the density of Au NPs, these procedures were repeated three times. The BiY$_2$Fe$_5$O$_{12}$ thin film was then formed by the MOD method[20]. The MOD solution for the Pt/BiY$_2$Fe$_5$O$_{12}$/Au-NP sample A (B) includes Bi, Y, and Fe carboxylates, dissolved in organic solvents with the concentration of 3 % (5 %). Its chemical composition is Bi:Y:Fe = 1:2:5. The solution for the sample A (B) was spin-coated on the Gd$_3$Ga$_5$O$_{12}$ substrate with Au NPs at 5000 r.p.m. for 60 seconds, followed by a drying step at 50 °C (170 °C) for 3 minutes and pre-annealing at 480 °C (550 °C) for 1 hour (5 minutes). To increase the thickness of the BiY$_2$Fe$_5$O$_{12}$ film, these processes (from spin-coating to pre-annealing) were repeated three times. Then, the sample A (B) was annealed at 725 °C (680 °C) for 15 hours in air to form a crystallized BiY$_2$Fe$_5$O$_{12}$ film. The cross-sectional transmission electron microscope images in Figs. 3e and 3j indicate that a crystalline BiY$_2$Fe$_5$O$_{12}$ film was formed on the Gd$_3$Ga$_5$O$_{12}$ substrate. Finally, a 5-nm-thick Pt film was deposited over the whole surface of the BiY$_2$Fe$_5$O$_{12}$ film by using an rf magnetron sputtering.

**Preparation process of Pt/Gd$_3$Ga$_5$O$_{12}$/Au-NP sample.** The Pt/Gd$_3$Ga$_5$O$_{12}$/Au-NP sample used in Fig. 2f consists of a Pt/Gd$_3$Ga$_5$O$_{12}$ bilayer film with Au NPs embedded in the Gd$_3$Ga$_5$O$_{12}$ layer. After forming Au NPs on a Gd$_3$Ga$_5$O$_{12}$ (111) substrate with the procedures described above, an additional Gd$_3$Ga$_5$O$_{12}$ thin film was coated on them by means of the MOD method. The MOD solution for the Gd$_3$Ga$_5$O$_{12}$ film includes Gd and Ga carboxylates, dissolved in organic solvents with the concentration of 3 %. Its chemical composition is Gd:Ga = 3:5. The spin-coating and annealing processes for the Gd$_3$Ga$_5$O$_{12}$ film are the same as those for the BiY$_2$Fe$_5$O$_{12}$ film in the Pt/BiY$_2$Fe$_5$O$_{12}$/Au-NP sample A. Finally, a 5-nm-thick Pt film was deposited over the whole surface of the Gd$_3$Ga$_5$O$_{12}$ film. The mean distance between the Pt film and Au NPs in the Pt/Gd$_3$Ga$_5$O$_{12}$/Au-NP sample is ~100 nm, which is comparable to that in the Pt/BiY$_2$Fe$_5$O$_{12}$/Au-NP sample.

**Acknowledgements**

The authors thank Y. Ohnuma, T. Kikkawa, S. Daimon, A. Kirihara, M. Ishida and B. Hillebrands for valuable discussions. This work was supported by PRESTO-JST 'Phase Interfaces for Highly Efficient Energy Utilization', CREST-JST 'Creation of Nanosystems with Novel Functions through Process Integration', Grant-in-Aid for Young Scientists (A) (25707029) from MEXT, Japan, Grant-in-Aid for Scientific Research (A) (24244051) from MEXT, Japan, LC-IMR of Tohoku University, the Sumitomo Foundation, the Tanikawa Fund Promotion of Thermal Technology, the Casio Science Promotion Foundation, and the Iwatani Naoji Foundation.

**Author contributions**

K.U. and E.S. planned and supervised the study. K.U. designed the experiments, prepared the samples, collected and analysed all of the data, and performed the numerical calculation. H.A. and S.M. developed the explanation of the experiments. D.K. performed the scanning electron microscopy. S.I. performed the transmission electron microscopy. Z.Q. supported the sample preparation. K.U. and H.A. wrote the manuscript with input from E.S. and S.M. All authors discussed the results.

**Additional information**

Correspondence and requests for materials should be addressed to K.U.

**Competing financial interests**

The authors declare no competing financial interests.

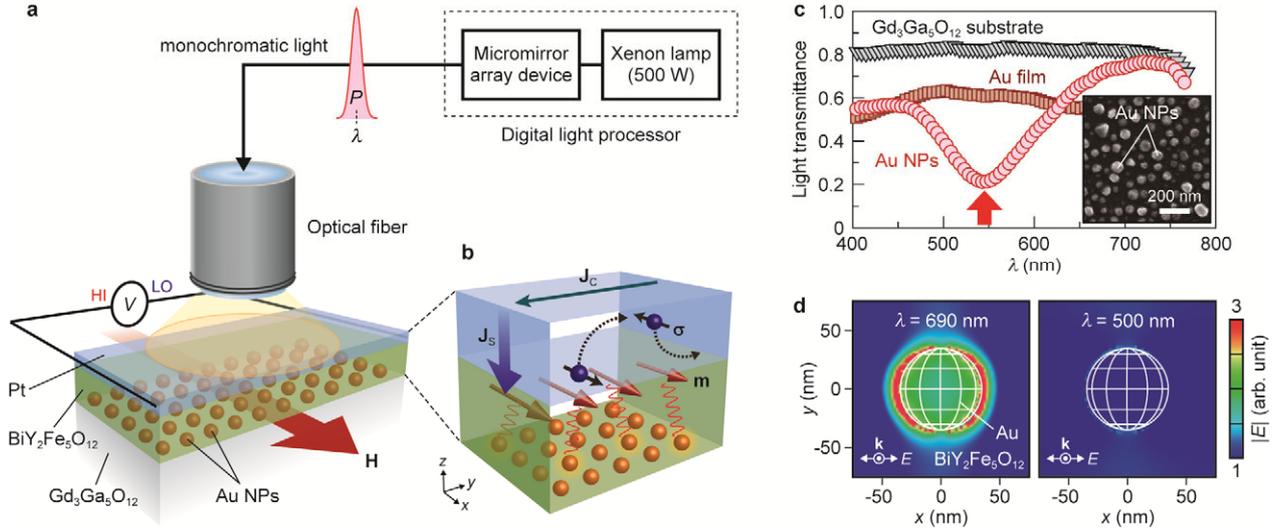

**Figure 1 | Experimental set-up. a**, Experimental configuration and sample structure for measuring the plasmonic spin pumping. The Pt/BiY$_2$Fe$_5$O$_{12}$/Au-nanoparticle (NP) sample is of a 10 × 5 mm$^2$ rectangular shape. The sample was illuminated with monochromatic light with the wavelength $\lambda$ (400-770 nm), power $P$, bandwidth of 20 nm, and spot diameter of ~5 mm from the Pt-film side at normal incidence by using a digital light processor comprising a 500 W Xenon lamp and a micromirror array device. **b**, Plasmonic spin pumping and inverse spin Hall effect (ISHE) in the Pt/BiY$_2$Fe$_5$O$_{12}$/Au-NP sample. **H**, **m**, **J**$_C$, **J**$_S$, and **σ** denote the external magnetic field vector (with magnitude $H$), magnetic moment vector, charge current generated by the ISHE, spatial direction of the spin current generated by the plasmonic spin pumping, and spin-polarization vector of the spin current, respectively. **c**, $\lambda$ dependence of the light transmittance of a 0.5-mm-thick Gd$_3$Ga$_5$O$_{12}$ substrate, a 15-nm-thick Au film, and Au NPs on the substrate. The light-transmittance spectrum of the Au-NP sample exhibits clear dip structure due to the SPR, while that of the Au film does not exhibit strong $\lambda$ dependence. The position of the SPR wavelength for the Au-NP sample is marked with a red arrow. The inset to **c** shows a scanning electron microscope image of Au NPs on the Gd$_3$Ga$_5$O$_{12}$ substrate, measured before forming the BiY$_2$Fe$_5$O$_{12}$ and Pt layers. **d**, Simulated distributions of the electric field intensity $|E|$ in a BiY$_2$Fe$_5$O$_{12}$/Au-NP model at $\lambda$ = 690 nm and 500 nm, calculated using a finite-difference time-domain method. These $|E|$ distributions are excited by plane electromagnetic waves with the polarization along the $x$ direction and the wave vector **k** along the $z$ direction (see SI section A1 for details).

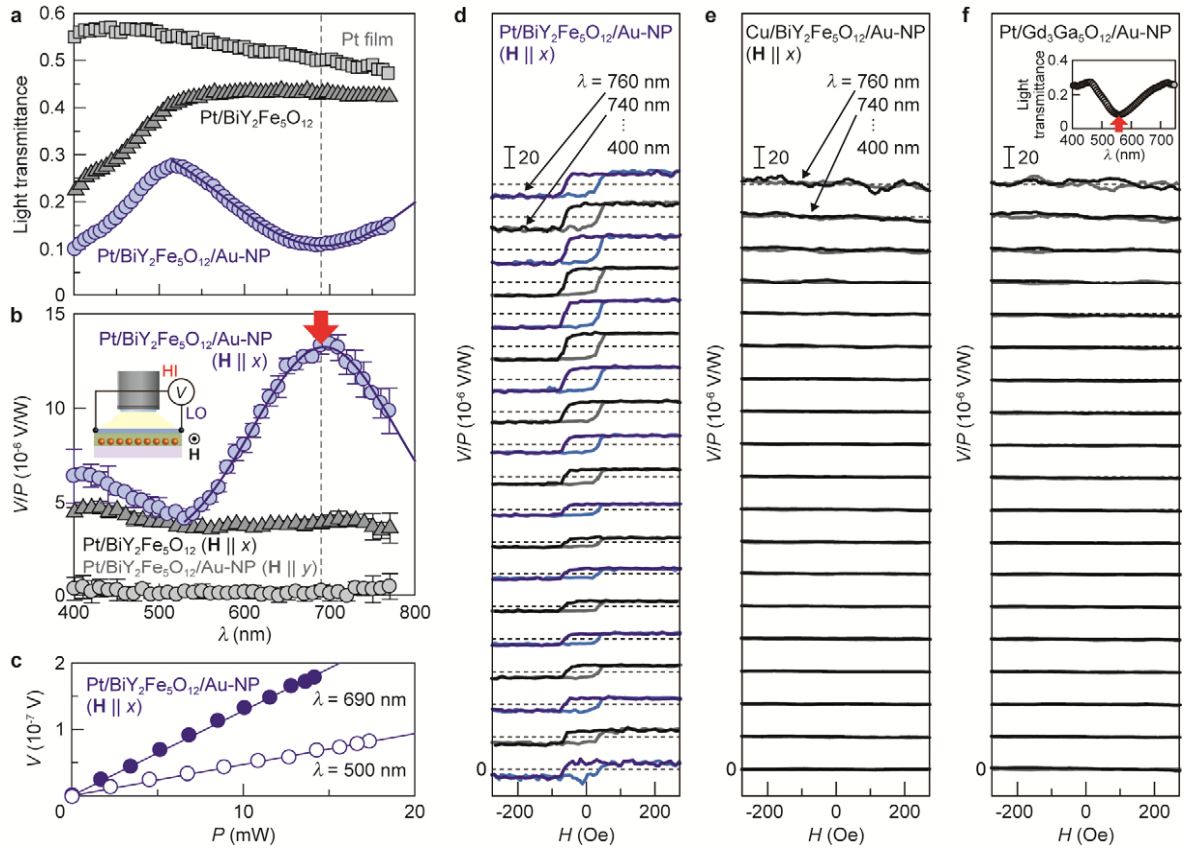

**Figure 2 | Observation of plasmonic spin pumping. a**, $\lambda$ dependence of the light transmittance of the Pt/BiY$_2$Fe$_5$O$_{12}$/Au-NP sample, the Pt/BiY$_2$Fe$_5$O$_{12}$ sample, and a 5-nm-thick Pt film on a Gd$_3$Ga$_5$O$_{12}$ substrate. The Pt/BiY$_2$Fe$_5$O$_{12}$/Au-NP and Pt/BiY$_2$Fe$_5$O$_{12}$ samples were prepared at the same time. In the Pt/BiY$_2$Fe$_5$O$_{12}$ sample, the Pt and BiY$_2$Fe$_5$O$_{12}$ films were formed directly on a Gd$_3$Ga$_5$O$_{12}$ substrate without Au NPs. **b**, $\lambda$ dependence of the electric voltage between the ends of the Pt layer normalized by the incident light power, $V/P$, in the Pt/BiY$_2$Fe$_5$O$_{12}$/Au-NP and Pt/BiY$_2$Fe$_5$O$_{12}$ samples, measured when **H** of $H = 200$ Oe was applied along the $x$ or $y$ direction. The in-plane coercive force of the BiY$_2$Fe$_5$O$_{12}$ layer is around 30 Oe; the magnetization of BiY$_2$Fe$_5$O$_{12}$ is aligned along the **H** direction at $H = 200$ Oe. The position of the SPR wavelength is marked with a red arrow. A blue solid curve in **a** (**b**) was obtained by fitting the observed dip (peak) structure using a Gaussian function. The error bars represent the standard deviation of the measurements. **c**, $P$ dependence of $V$ in the Pt/BiY$_2$Fe$_5$O$_{12}$/Au-NP sample at $\lambda = 690$ nm and 500 nm, measured when **H** of $H = 200$ Oe was applied along the $x$ direction. **d-f**, $H$ dependence of $V/P$ in the Pt/BiY$_2$Fe$_5$O$_{12}$/Au-NP (**d**), Cu/BiY$_2$Fe$_5$O$_{12}$/Au-NP (**e**), and Pt/Gd$_3$Ga$_5$O$_{12}$/Au-NP (**f**) samples for various values of $\lambda$, measured when **H** was along the $x$ direction. The inset to **f** shows the $\lambda$ dependence of the light transmittance of the Pt/Gd$_3$Ga$_5$O$_{12}$/Au-NP sample. All the control experiments were performed by using the samples of a $10 \times 5$ mm$^2$ rectangular shape.

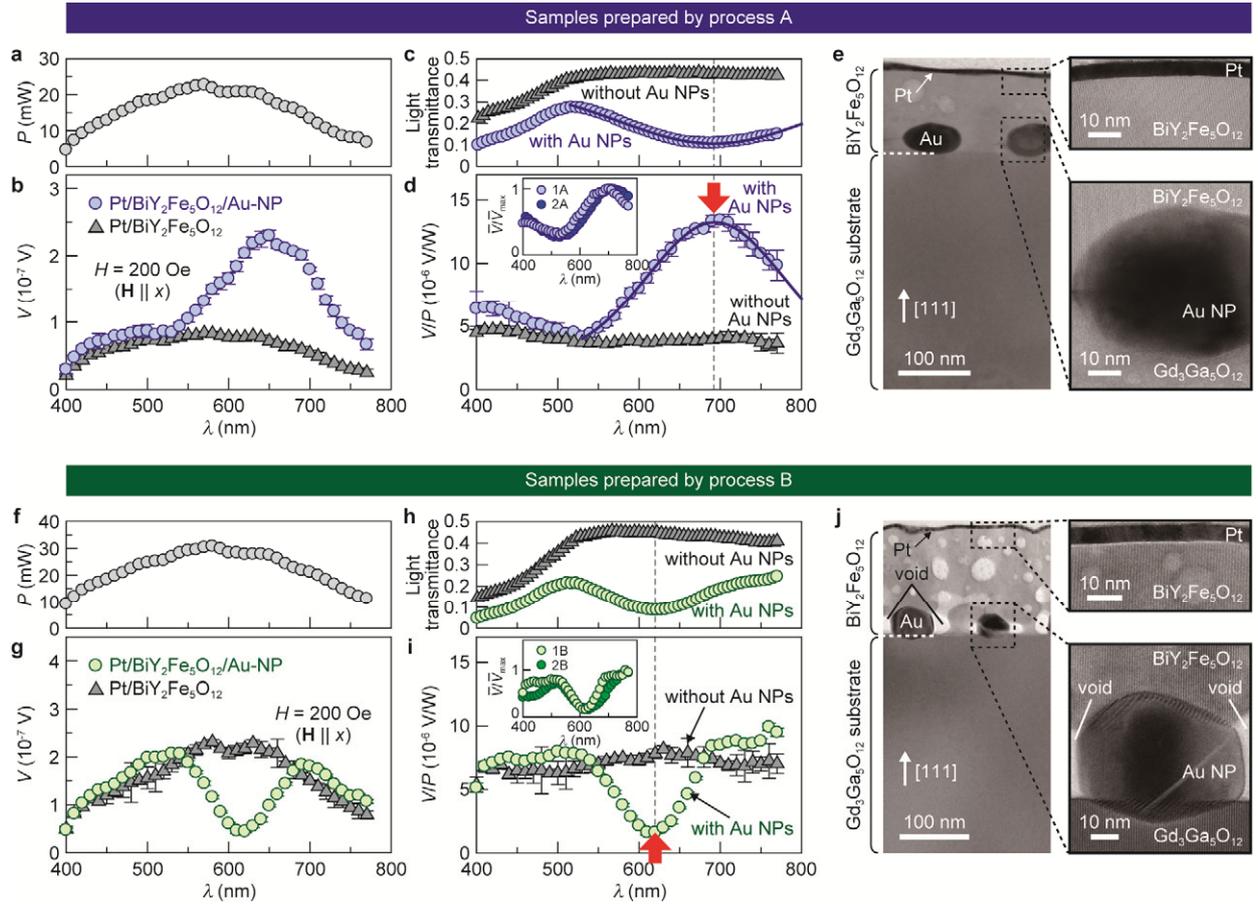

**Figure 3 | Evidence for magnon excitation by surface plasmons. a**, $\lambda$ dependence of $P$, applied when the data in **b** was measured. **b**, $\lambda$ dependence of $V$ in the Pt/BiY$_2$Fe$_5$O$_{12}$/Au-NP and Pt/BiY$_2$Fe$_5$O$_{12}$ samples prepared by the process A, measured when **H** of $H$ = 200 Oe was applied along the $x$ direction. The error bars represent the standard deviation of the measurements. **c**, $\lambda$ dependence of the light transmittance of the Pt/BiY$_2$Fe$_5$O$_{12}$/Au-NP and the Pt/BiY$_2$Fe$_5$O$_{12}$ samples prepared by the process A. **d**, $\lambda$ dependence of $V/P$ in the Pt/BiY$_2$Fe$_5$O$_{12}$/Au-NP and the Pt/BiY$_2$Fe$_5$O$_{12}$ samples prepared by the process A. The inset to **d** shows the $\lambda$ dependence of $\bar{V}/\bar{V}_{max}$ in two different Pt/BiY$_2$Fe$_5$O$_{12}$/Au-NP samples 1A and 2A, where $\bar{V} \equiv V/P$ and $\bar{V}_{max}$ is the maximum value of the $\lambda$-$\bar{V}$ spectrum. **e**, Cross-sectional transmission electron microscope images of the Pt/BiY$_2$Fe$_5$O$_{12}$/Au-NP sample prepared by the process A. **f-j**, Experimental results for the the Pt/BiY$_2$Fe$_5$O$_{12}$/Au-NP and Pt/BiY$_2$Fe$_5$O$_{12}$ samples prepared by the process B. The inset to **i** shows the $\lambda$ dependence of $\bar{V}/\bar{V}_{max}$ in two different Pt/BiY$_2$Fe$_5$O$_{12}$/Au-NP samples 1B and 2B. The Pt/BiY$_2$Fe$_5$O$_{12}$/Au-NP sample B has small voids with the size of 20-30 nm between the BiY$_2$Fe$_5$O$_{12}$ film and Au NPs, while Au NPs in the sample A are densely embedded in BiY$_2$Fe$_5$O$_{12}$. The difference in the SPR wavelength between the Pt/BiY$_2$Fe$_5$O$_{12}$/Au-NP samples A and B is attributed to the difference in the dielectric constant of the medium around Au NPs; since Au NPs in the Pt/BiY$_2$Fe$_5$O$_{12}$/Au-NP sample B are surrounded by voids, the SPR wavelength of the sample B is closer to that of plain Au NPs without BiY$_2$Fe$_5$O$_{12}$ (compare Figs. 1c, 3c, and 3h).

## A. Numerical calculation

### A1. Methods for FDTD simulation in Fig. 1d

To investigate the electromagnetic field distributions induced by surface plasmons, we performed the numerical calculation based on a finite-difference time-domain (FDTD) method[23] using KeyFDTD (Science Technology Research Institute, Japan). The simulation results shown in Fig. 1d were obtained from a model comprising a $BiY_2Fe_5O_{12}$ rectangular parallelepiped with the size of $150 \times 150 \times 140$ nm$^3$ and a Au spheroid with the in-plane diameter $d_{Au}$ of 70 nm and the height $h_{Au}$ of 40 nm embedded at the centre of the $BiY_2Fe_5O_{12}$, where the distance $L$ between the top of the $BiY_2Fe_5O_{12}$ and the top of the Au is 50 nm (Fig. S1a). The unit cell of the model is $2 \times 2 \times 1$ nm$^3$. A flat light source is placed on the top $x$-$y$ plane of the $BiY_2Fe_5O_{12}$ rectangular parallelepiped, which generates plane electromagnetic waves with the polarization along the $x$ direction and the wave vector **k** along the $z$ direction (note that the incident light used in the experiments is unpolarized). Here, the $x$, $y$, and $z$ directions are defined in Fig. S1a. At the outer boundaries of the $BiY_2Fe_5O_{12}$, we set a periodic boundary condition. The refraction-index and absorption-constant spectra of $BiY_2Fe_5O_{12}$ and Au in the calculations were obtained from refs. 31 and 32, respectively. Under this condition, we calculated the distribution of the electric field intensity $|E|$ while fixing the wavelength of the incident electromagnetic waves. In Fig. 1d, we plot the $|E|$ distribution in the $BiY_2Fe_5O_{12}$/Au-nanoparticle (NP) model in the $x$-$y$ plane across the centre of the Au spheroid.

### A2. Electromagnetic coupling between Pt film and Au nanoparticles

Here, we investigate the electromagnetic coupling between the Pt film and Au NPs in the Pt/$BiY_2Fe_5O_{12}$/Au-NP sample, which might affect the voltage signal under the surface-plasmon-resonance (SPR) condition. To do this, we compare the simulated distributions of the electric field intensity $|E|$ induced by the SPR between the $BiY_2Fe_5O_{12}$/Au-NP models with and without the Pt layer. The $BiY_2Fe_5O_{12}$/Au-NP model used here is the same as that described in the subsection A1 except for the values of $L$, $d_{Au}$, and $h_{Au}$ (Fig. S1a). The Pt/$BiY_2Fe_5O_{12}$/Au-NP model consists of the $BiY_2Fe_5O_{12}$/Au-NP model and a Pt rectangular parallelepiped with the size of $150 \times 150 \times 5$ nm$^3$ attached on the top of the $BiY_2Fe_5O_{12}$ (Fig. S1b). The wavelength of the incident electromagnetic waves is fixed at $\lambda = 690$ nm. The refraction-index and absorption-constant spectra of Pt in the calculations were obtained from ref. 33.

Figures S1c and S1d respectively show the $|E|$ distributions in the $BiY_2Fe_5O_{12}$/Au-NP and Pt/$BiY_2Fe_5O_{12}$/Au-NP models in the $z$-$x$ plane across the centre of the Au spheroid for various values of $L$, simulated when $d_{Au} = 70$ nm and $h_{Au} = 40$ nm. In both the models, strong $|E|$ is induced in the vicinity of the $BiY_2Fe_5O_{12}$/Au-NP interface due to the localized SPR. Importantly, the $|E|$ distributions in $BiY_2Fe_5O_{12}$ are not affected by the presence of Pt irrespective of the distance $L$, indicating that the electromagnetic coupling between the Pt film and Au NPs is very weak and no SPR mode due to the coupling appears in the configuration used in this study (note that the values of $L$ used in the calculations are smaller than the mean distance between the Pt film and Au NPs in the experiments). We also calculated the $|E|$ distributions in the $BiY_2Fe_5O_{12}$/Au-NP and Pt/$BiY_2Fe_5O_{12}$/Au-NP models for various values of $d_{Au}$ and $h_{Au}$ (Fig. S2). The simulation results confirm again that, although the $|E|$ distributions in $BiY_2Fe_5O_{12}$ under the SPR condition depend on the size of the Au NP, they are not affected by the presence of Pt irrespective of the values of $d_{Au}$ and $h_{Au}$. These results show that the temperature rise of the Pt layer does not change by the SPR in Au NPs, supporting our interpretation that the voltage enhancement in the Pt/$BiY_2Fe_5O_{12}$/Au-NP sample A under the SPR condition (Fig. 2b) is attributed to the plasmonic spin pumping and irrelevant to the heating of the sample.

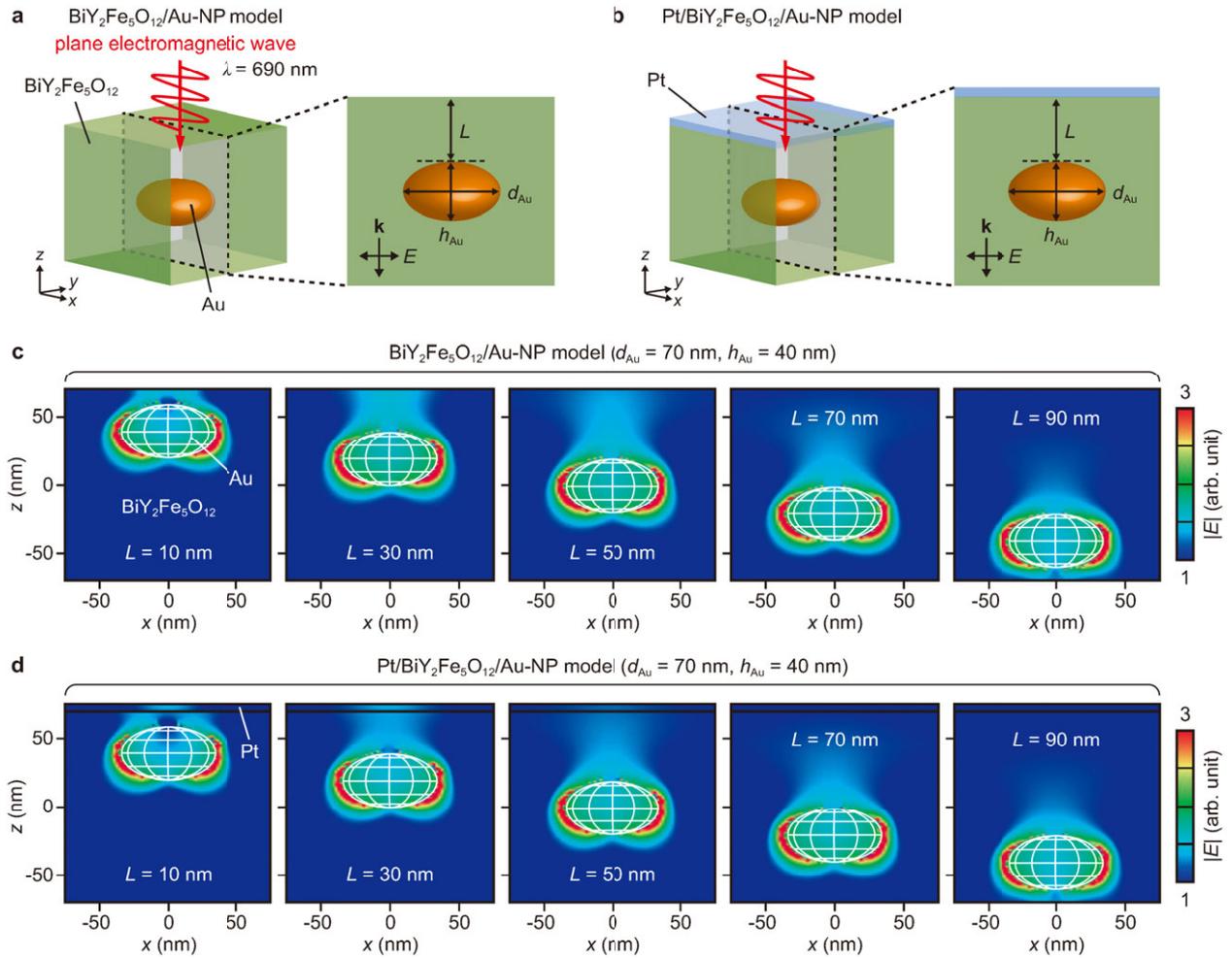

**Figure S1 | Dependence of electric field distribution on the position of Au nanoparticles.** **a,b,** Schematic illustrations of the $BiY_2Fe_5O_{12}$/Au-NP (**a**) and Pt/$BiY_2Fe_5O_{12}$/Au-NP (**b**) models used for the FDTD simulation. $L$ denotes the distance between the top of the $BiY_2Fe_5O_{12}$ rectangular parallelepiped and the top of the Au spheroid. $d_{Au}$ and $h_{Au}$ denote the in-plane diameter and height of the Au, respectively. **c,d,** Simulated distributions of the electric field intensity $|E|$ in the $BiY_2Fe_5O_{12}$/Au-NP (**c**) and Pt/$BiY_2Fe_5O_{12}$/Au-NP (**d**) models at $\lambda = 690$ nm in the $z$-$x$ plane across the centre of the Au spheroid for various values of $L$, calculated when $d_{Au} = 70$ nm and $h_{Au} = 40$ nm.

### A3. Effect of voids on electric field and power-dissipation distributions

In this subsection, we compare the simulated distributions of the electric field intensity $|E|$ induced by the SPR between the Pt/$BiY_2Fe_5O_{12}$/Au-NP models with and without a void. To model voids, a vacuum spheroid with the in-plane diameter of 100 nm and the height of 40 nm is added to the Pt/$BiY_2Fe_5O_{12}$/Au-NP model around the Au spheroid (Fig. S3d), where $L = 70$ nm, $d_{Au} = 70$ nm, and $h_{Au} = 40$ nm. According to the experimental results in Fig. 3, the wavelength of the incident electromagnetic waves is fixed at $\lambda = 690$ nm ($\lambda = 630$ nm) for the Pt/$BiY_2Fe_5O_{12}$/Au-NP model with (without) the void.

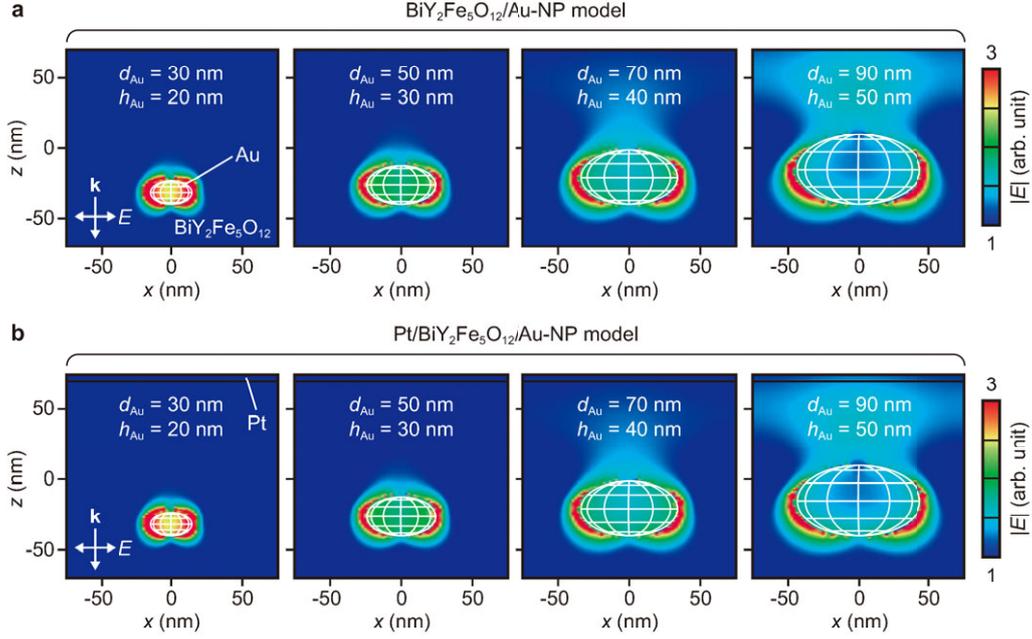

**Figure S2 | Dependence of electric field distribution on the size of Au nanoparticles. a,b,** Simulated $|E|$ distributions in the $BiY_2Fe_5O_{12}$/Au-NP (**a**) and Pt/$BiY_2Fe_5O_{12}$/Au-NP (**b**) models at $\lambda = 690$ nm in the $z$-$x$ plane across the centre of the Au spheroid for various values of $d_{Au}$ and $h_{Au}$.

In Figs. S3b and S3e, we show the $|E|$ distributions in the Pt/$BiY_2Fe_5O_{12}$/Au-NP models with and without the void, respectively. As shown in the previous results, in the model without the void, the strong near fields are induced in $BiY_2Fe_5O_{12}$ in the vicinity of the Au spheroid due to the SPR. In contrast, in the model with the void, most of the photon energy is confined in the void, indicating that the plasmon-induced near fields cannot interact with magnons in $BiY_2Fe_5O_{12}$ with voids. These simulation results are consistent with our experiments in Fig. 3, where the plasmonic spin pumping appears in the Pt/$BiY_2Fe_5O_{12}$/Au-NP sample A (without voids) while only heating effect appears in the sample B (with voids).

Next we show the calculated power dissipation $Q$ under the SPR condition in the same Pt/$BiY_2Fe_5O_{12}$/Au-NP models with and without the void. As shown in Figs. S3c and S3f, the maximum of $Q$ is in the vicinity of the Au spheroid in both the models, indicating that the heating due to the SPR is induced near Au NPs irrespective of the presence of voids. These simulation results allow us to conclude again that the temperature gradient induced by the SPR is of an opposite sign to that induced by the heating of the Pt layer in both the Pt/$BiY_2Fe_5O_{12}$/Au-NP samples A and B.

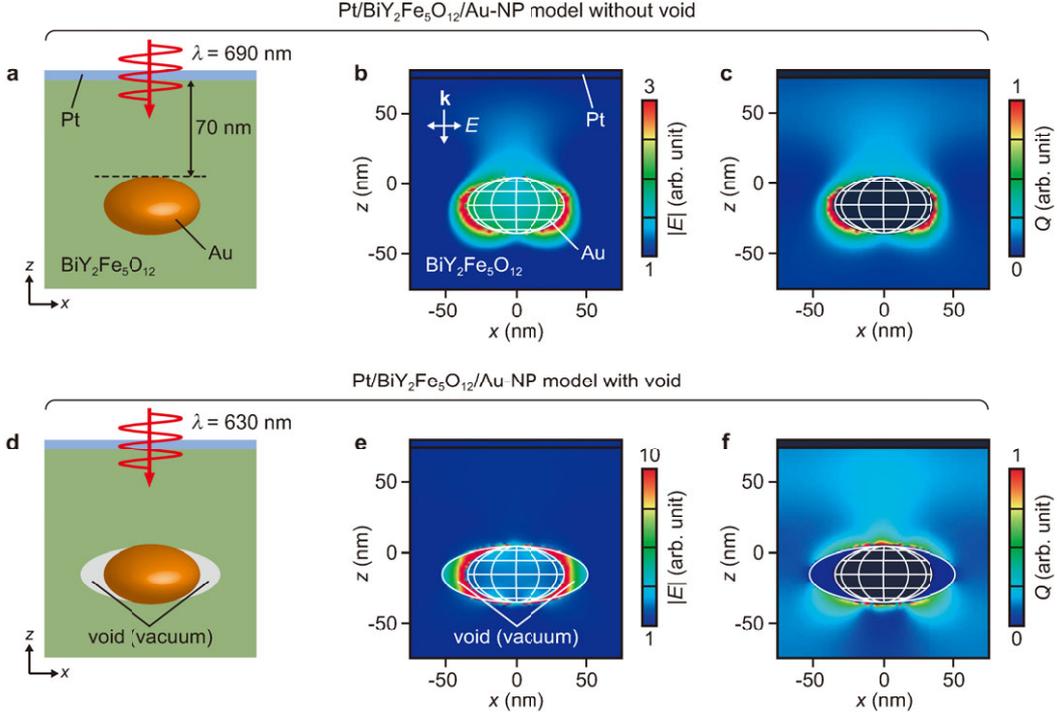

**Figure S3 | Effect of voids. a**, Schematic illustration of the $Pt/BiY_2Fe_5O_{12}/Au$-NP models without the void used for the FDTD simulation. **b,c**, Simulated distributions of $|E|$ (**b**) and the power dissipation $Q$ (**c**) in the $Pt/BiY_2Fe_5O_{12}/Au$-NP models without the void at $\lambda = 690$ nm in the $z$-$x$ plane across the centre of the Au spheroid. **d-f**, Schematic illustration and simulation results for the $Pt/BiY_2Fe_5O_{12}/Au$-NP models with the void at $\lambda = 630$ nm.

## B. Detection of plasmonic spin pumping using Au film

In the main text, we report the observation of the plasmonic spin pumping in the $Pt/BiY_2Fe_5O_{12}/Au$-NP sample. In this section, we show that the plasmonic spin pumping appears also in a $Au/BiY_2Fe_5O_{12}/Au$-NP sample, where the Pt film is replaced with a 5-nm-thick Au film (Fig. S4b). Since Au is a typical metal far from ferromagnetism[34,35], the observation of the plasmonic spin pumping in this sample allows us to conclude that the plasmonic spin pumping cannot be explained by extrinsic artifacts induced by the magnetic proximity[21,26] at the $Pt/BiY_2Fe_5O_{12}$ interface and the static ferromagnetism in Pt. The measurements were performed in the configuration shown in Fig. 1a, where the $Au/BiY_2Fe_5O_{12}/Au$-NP sample was illuminated with monochromatic light with the wavelength $\lambda$, power $P$, bandwidth of 50 nm, and spot diameter of $\sim 5$ mm from the Au-film side at normal incidence by using a digital light processor comprising 500 W Xenon lamp and micromirror array device (Gooch & Housego, OL490).

Figure S4a shows the $\lambda$ dependence of the electric voltage between the ends of the Au layer normalized by the incident light power, $V/P$, in the $Au/BiY_2Fe_5O_{12}/Au$-NP and $Au/BiY_2Fe_5O_{12}$ samples, measured when the in-plane magnetic field of 200 Oe was applied perpendicular to the inter-electrode direction ($x$ direction). We found that, also in the $Au/BiY_2Fe_5O_{12}/Au$-NP sample, the finite voltage signal appears in the Au layer at all the $\lambda$ values and the magnitude of $V/P$ is enhanced under the SPR condition (compare Figs. S4a and S4c). This behaviour is very similar to that of the $V/P$ signal observed in the Pt/BiY$_2$Fe$_5$O$_{12}$/Au-NP sample (Fig. 2b), confirming that the plasmonic spin pumping appears even in the absence of the static ferromagnetism induced by the magnetic proximity.

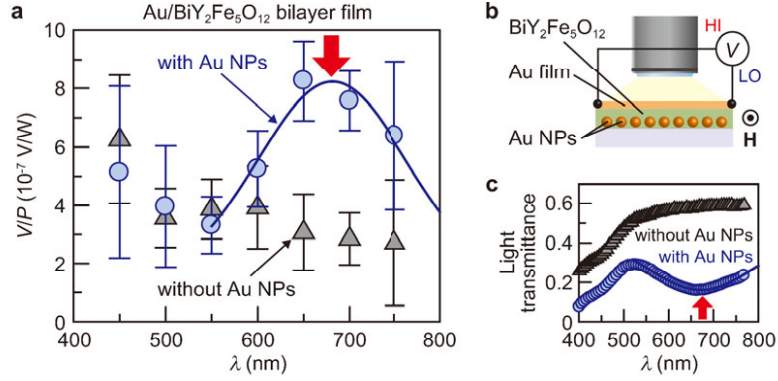

**Figure S4 | Plasmonic spin pumping in Au/BiY$_2$Fe$_5$O$_{12}$/Au-nanoparticle system. a**, $\lambda$ dependence of $V/P$ in the Au/BiY$_2$Fe$_5$O$_{12}$/Au-NP and Au/BiY$_2$Fe$_5$O$_{12}$ samples, measured when **H** of $H = 200$ Oe was applied along the $x$ direction. The BiY$_2$Fe$_5$O$_{12}$ layers of the samples were prepared by the process A (see Methods in the main text). **b**, Experimental configuration and sample structure for measuring the plasmonic spin pumping in the Au/BiY$_2$Fe$_5$O$_{12}$/Au-NP sample. **c**, $\lambda$ dependence of the light transmittance of the Au/BiY$_2$Fe$_5$O$_{12}$/Au-NP and Au/BiY$_2$Fe$_5$O$_{12}$ samples.

## C. Voltage measurements using thicker Pt films

In this section, we report on the voltage measurements in the Pt/BiY$_2$Fe$_5$O$_{12}$/Au-NP and Pt/BiY$_2$Fe$_5$O$_{12}$ samples in which the 5-nm-thick Pt films are replaced with 15-nm-thick Pt films. Due to the thicker Pt films, the light transmittance of these samples is much smaller than that of the samples used in the main text (compare Figs. 2a and S5a). The voltage measurements were also performed in the configuration shown in Fig. 1a.

The gray triangle data points in Fig. S5b show the $\lambda$ dependence $V/P$ in the Pt(15 nm)/BiY$_2$Fe$_5$O$_{12}$ sample without Au NPs. We observed finite spin-current signals at all the $\lambda$ values in this sample and found that its $\lambda$ dependence is similar to the data in the Pt(5 nm)/BiY$_2$Fe$_5$O$_{12}$ sample without Au NPs (compare Figs. 2b and S5b). Furthermore, we also checked that similar signals appear even when the surface of the Pt(15 nm)/BiY$_2$Fe$_5$O$_{12}$ sample is coated with black ink (Figs. S5b and S5c), which almost completely blocks the transmission of light and the interaction between photons and magnons in BiY$_2$Fe$_5$O$_{12}$ (Fig. S5a). These results confirm again that the background spin-current signals in the Pt/BiY$_2$Fe$_5$O$_{12}$ samples without Au NPs are attributed to the heating of the samples, or the spin Seebeck effect (note that the enhancement of the heating signal in the Black-ink/Pt(15 nm)/BiY$_2$Fe$_5$O$_{12}$ sample is due to the larger temperature rise caused by light absorption by the black ink).

The blue circle data points in Fig. S5b show the $\lambda$ dependence $V/P$ in the Pt(15 nm)/BiY$_2$Fe$_5$O$_{12}$ sample with Au NPs. Also in this sample, we observed the voltage enhancement due to the plasmonic spin pumping under the SPR condition (compare the $V/P$ spectra in the samples with and without Au NPs in Fig. S5b). However, the voltage enhancement observed here is weaker than that in the Pt(5 nm)/BiY$_2$Fe$_5$O$_{12}$/Au-NP sample shown in Fig. 2b. This is because the thicker Pt layer blocks the transmission of light and reduces the magnitude of the near fields in the BiY$_2$Fe$_5$O$_{12}$ layer, a situation which changes the relative magnitude between the plasmonic-spin-pumping signal and the background spin Seebeck signal.

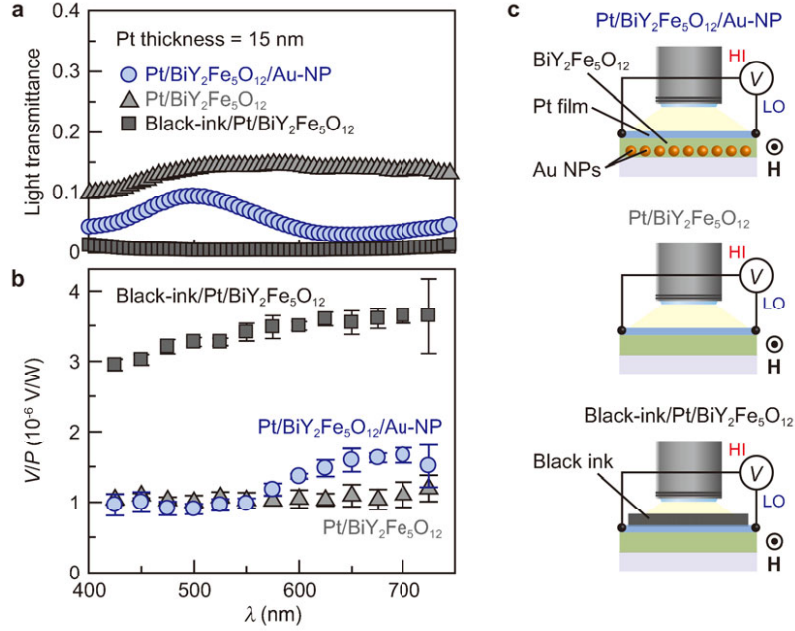

**Figure S5** | **Voltage measurements using thicker Pt films.** **a**, $\lambda$ dependence of the light transmittance of the Pt(15 nm)/BiY$_2$Fe$_5$O$_{12}$/Au-NP, Pt(15 nm)/BiY$_2$Fe$_5$O$_{12}$, and Black-ink/Pt(15 nm)/BiY$_2$Fe$_5$O$_{12}$ samples. The BiY$_2$Fe$_5$O$_{12}$ layers of the samples were prepared by the process A (see Methods in the main text). **b**, $\lambda$ dependence of $V/P$, measured when **H** of $H = 200$ Oe was applied along the $x$ direction. **c**, Experimental configuration and sample structure for measuring voltage in the Pt(15 nm)/BiY$_2$Fe$_5$O$_{12}$/Au-NP, Pt(15 nm)/BiY$_2$Fe$_5$O$_{12}$, and Black-ink/Pt(15 nm)/BiY$_2$Fe$_5$O$_{12}$ samples.

## D. Linear response theory of plasmonic spin pumping

In this section, we formulate a linear response theory of the plasmonic spin pumping and show that the experimental observation can be explained as a consequence of the energy transfer from plasmon-induced near-field photons to magnons in the BiY$_2$Fe$_5$O$_{12}$ layer. Interaction of photons and magnons has been extensively studied in the context of Raman scattering by magnons[27-30]. From these studies, it has been established that the dominant coupling mechanism is given by an electric-dipole coupling via a spin-orbit interaction. Because the energy of incident near-field photons (1.6-3.1 eV) is much higher than the energy scale of magnons ($< 0.1$ eV), a linear coupling between photons and magnons via a direct magnetic-dipole interaction[36], i.e., optical generation of magnons by light absorption, is irrelevant to the present discussion, as discussed in ref. 30.

Following ref. 37, the Hamiltonian that gives rise to the scattering of light by a magnetic system is written as

$$H' = \sum_{j_1,j_2} \sum_{r_i} E_1^{j_1}(r_i) E_2^{j_2}(r_i) \Pi^{j_1,j_2}(r_i), \tag{S1}$$

where $E_{1(2)}$ is the incident (scattered) electric field at position $r_i$, $j_{1(2)}$ represents a Cartesian component of the field vector, and $\Pi^{j_{1(2)}}(r_i)$, a term containing the spin operator $S(r_i)$, describes the spin-dependent polarizability tensor at $r_i$. For systems such as $BiY_2Fe_5O_{12}$, equation (S1) can be expressed in the form[38]

$$H' = \frac{v_0}{2} \sum_{r_i} \left( \left[ \Gamma^- E_1^z(r_i) E_2^-(r_i) + \Gamma^+ E_1^-(r_i) E_2^z(r_i) \right] S^-(r_i) + \left[ \Gamma^+ E_1^z(r_i) E_2^+(r_i) + \Gamma^- E_1^-(r_i) E_2^z(r_i) \right] S^+(r_i) \right) \tag{S2}$$

to the lowest order in the spin operators, where $S^\pm = S^x \pm iS^y$, $E^\pm = E^x \pm iE^y$, and $v_0 = V/N_0$ is the effective block spin volume with the number of unit cells $N_0$. In the above equation, $\Gamma^\pm = G \pm iM$ is the dimensionless magnon-photon coupling constant, where we assume that $G$ and $M$ are real numbers. Note that the magnon-photon coupling in equation (S2) forms the basis for the Brillouin light scattering investigation of $Y_3Fe_5O_{12}$ (ref. 39). The incident electric field $E_1$ is an external c-number field with the representation

$$E_1(r,t) = \sum_{\zeta_1} \sum_{K_1=\pm K_0} E_{K_1,\zeta_1}(t) e^{iK_1 \cdot r}, \tag{S3}$$

where the coefficient satisfies $E_{-K_1,\zeta_1}(t) = E^*_{K_1,\zeta_1}(t)$ and is expressed as $E_{K_1,\zeta_1}(t) = \xi(K_1,\zeta_1) \mathcal{E}_{K_1,\zeta_1}$ with the wavenumber $K_1$, polarization $\zeta_1$, and amplitude $\mathcal{E}_{K_1,\zeta_1}$. The scattered electric field $E_2$ is, on the other hand, expressed in a quantized form[37]. It is custom to express this field by using the vector potential $A_2$ as $E_2(r,t) = -c^{-1} \partial_t A_2(r,t)$, where $c$ is the velocity of light. The vector potential is represented as

$$A_2(r,t) = \sum_{\zeta_2} \sum_{K_2} A_{K_2,\zeta_2}(t) e^{iK_2 \cdot r}, \tag{S4}$$

where $A_{K_2,\zeta_2}(t) = \sqrt{\frac{2\pi\hbar}{\nu_{K_2} V}} \xi(K_2,\zeta_2) [a_{K_2,\zeta_2}(t) + a^\dagger_{-K_2,\zeta_2}(t)]$ with the photon frequency $\nu_K = cK$, the photon annihilation and creation operators $a_{K_2,\zeta_2}$ and $a^\dagger_{-K_2,\zeta_2}$, and the polarization vector $\xi(K_2,\zeta_2)$. Using the linear spin-wave approximation

$$S^+(r_i) = \sqrt{\frac{2S_0}{N_0}} \sum_q b_q e^{iq \cdot r} \tag{S5}$$

and

$$S^-(r_i) = \sqrt{\frac{2S_0}{N_0}} \sum_q b^\dagger_q e^{-iq \cdot r} \tag{S6}$$

with the size of the localized spin $S_0$, the magnon-photon interaction is represented as

$$H' = -\frac{1}{c}\sqrt{\frac{S_0 v_0}{2}} \sum_q \sum_{K_1=\pm K_0} \left( \Gamma^- E^z_{-K_1} (\partial_t A^-_{K_1-q}) + \Gamma^+ E^-_{-K_1} (\partial_t A^z_{K_1-q}) \right) b_q + H.c., \tag{S7}$$

where H.c. means the Hermitian conjugate. In the present situation, the incident electric field $E_1$ comes from the near-field photons induced by the localized SPR in the Au NPs embedded in the $BiY_2Fe_5O_{12}$ film. Therefore, we assume that its polarization direction as well as its propagating direction should be averaged

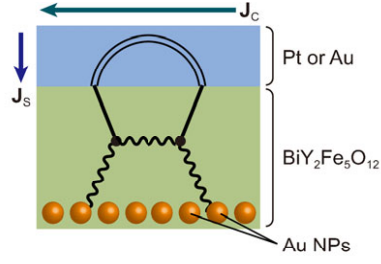

**Figure S6 | Feynman diagram for calculating the spin current generated by the plasmonic spin pumping.** The wavy lines represent a (near-field) photon propagator or external photon fields. The double line and bold lines represent itinerant-spin-density and magnon propagators, respectively.

out in the final step of our calculation.

Following the formalism developed in ref. 40, we now calculate the spin current $J_S$ generated by the plasmonic spin pumping. We consider a model shown in Fig. S6, and investigate the spin injection from a ferrimagnetic insulator (FI, in the experiments $BiY_2Fe_5O_{12}$) into an attached paramagnetic metal (PM, in the experiments Pt or Au) through the $s$-$d$ exchange interaction acting at the PM/FI interface:

$$\mathcal{H}_{sd} = J_{sd} \sum_{\bm{r}_0 \in \text{interface}} \bm{s}(\bm{r}_0) \cdot \bm{S}(\bm{r}_0), \tag{S8}$$

where $\bm{s}$ is the conduction-electrons' spin density in PM, $\bm{S}$ is the localized spin in FI, and $J_{sd}$ is the strength of the interface $s$-$d$ exchange coupling. The spin current $J_S$ generated in PM can be calculated as a rate of change of the spin density in PM as $J_S = \frac{\hbar}{2} \sum_{\bm{r}_i \in \text{PM}} \langle \partial_t \bm{s}(\bm{r}_i, t) \rangle$, where $\langle \cdots \rangle$ denotes the statistical average at a given time $t$. Assuming that the spin-orbit interaction is weak enough in the neighborhoods of the interface, the Heisenberg equation of motion for $\bm{s}$ yields

$$J_S = \sum_{\bm{q},\bm{k}} \frac{-2\mathcal{J}_{sd}^{\bm{k}-\bm{q}}\sqrt{S_0}}{\sqrt{2N_P N_F}} \int_{-\infty}^{\infty} \frac{d\omega}{2\pi} \text{Re} C^<_{\bm{k},\bm{q}}(\omega), \tag{S9}$$

where $\mathcal{J}_{sd}^{\bm{k}-\bm{q}}$ is the Fourier transform of $\mathcal{J}_{sd}(\bm{r}) = J_{sd} \sum_{\bm{r}_0 \in \text{interface}} a_S^3 \delta(\bm{r}-\bm{r}_0)$, $N_{P(F)}$ is the number of lattice sites in PM (FI), and we represent the effective block spin volume as $v_0 = a_S^3$ by introducing $a_S$. Here, $C^<_{\bm{k},\bm{q}}(\omega)$ is the Fourier transform of the interface correlation $C^<_{\bm{k},\bm{q}}(t,t') = -i\langle b_{\bm{q}}^+(t') s_{\bm{k}}^-(t) \rangle$ between the magnons and the spin density $s_{\bm{k}}^- = \frac{1}{2\sqrt{N_P}} \sum_{\bm{r}_i} [s^x(\bm{r}_i) - is^y(\bm{r}_i)] e^{-i\bm{k}\cdot\bm{r}_i}$.

The process relevant to the plasmonic spin pumping is shown in Fig. S6; in the case of the plasmon-induced spin injection, the near-field photons concomitant with surface plasmons excite only magnons in the $BiY_2Fe_5O_{12}$ film since they are localized in the vicinity of the $BiY_2Fe_5O_{12}$/Au-NP interface (Figs. 1d and S1-S3). Due to the similarity between this diagram and that for the acoustic spin pumping (ASP)[41,42] (Fig. 10 in ref. 40), the present calculation for the plasmonic spin pumping is mostly the same as that for the ASP, with the replacement of external *phonon* lines by external *photon* lines. One big difference is that the intermediate state is given by scattered photons in the present situation whereas it is given by magnons for the ASP. This is because the energy of external photons (1.6-3.1 eV) is much larger than that of magnons ($< 0.1$ eV), and most of the external photon energy contributing to this process is transferred to scattered photons, leaving a small energy transfer to magnons. Using the same procedure as in ref. 40, the spin current generated by the process shown in Fig. S6 is calculated to be

$$J_S = \frac{\sqrt{2} N_{\text{int}} J_{\text{sd}}^2 S_0^2}{2\hbar^3 N_P N_F} \left( v_0 |\mathcal{E}_{\boldsymbol{K}_0}|^2 \right) \sum_{\boldsymbol{k},\boldsymbol{q}} \Gamma^+ \Gamma^- B_{\boldsymbol{k},\boldsymbol{q}}(\nu_{\boldsymbol{K}_0}), \tag{S10}$$

where $N_{\text{int}}$ is the number of localized spins in FI at the interface. The quantity $B_{\boldsymbol{k},\boldsymbol{q}}(\nu_{\boldsymbol{K}_0})$ is defined by

$$B_{\boldsymbol{k},\boldsymbol{q}}(\nu_{\boldsymbol{K}_0}) = \left(\frac{\nu_{\boldsymbol{K}_0}}{c}\right)^2 \int_\omega \text{Im}\chi^R_{\boldsymbol{k}}(\omega) \text{Im} D^R_{\boldsymbol{q}-\boldsymbol{K}_0}(\omega - \nu_{\boldsymbol{K}_0}) |X^R_{\boldsymbol{q}}(\omega)|^2 \left[\coth\left(\frac{\hbar(\omega - \nu_{\boldsymbol{K}_0})}{2k_B T}\right) - \coth\left(\frac{\hbar\omega}{2k_B T}\right)\right]$$
$$+ (\nu_{\boldsymbol{K}_0} \to -\nu_{\boldsymbol{K}_0}; \boldsymbol{K}_0 \to -\boldsymbol{K}_0) \tag{S11}$$

with the shorthand notation $\int_\omega = \int_{-\infty}^{\infty} \frac{d\omega}{2\pi}$. In the above equation, $\chi^R_{\boldsymbol{k}}(\omega) = \chi_P/(1 + \lambda_{\text{sf}}^2 k^2 - i\omega\tau_{\text{sf}})$ is the retarded component of the itinerant-spin-density propagator in PM with $\chi_P$, $\lambda_{\text{sf}}$, and $\tau_{\text{sf}}$ being respectively the paramagnetic susceptibility, spin diffusion length, and spin relaxation time. Also, $X^R_{\boldsymbol{q}}(\omega) = (\omega - \widetilde{\omega}_{\boldsymbol{q}} + i\alpha\omega)^{-1}$ is the retarded component of the magnon propagator with $\widetilde{\omega}_{\boldsymbol{q}} = \gamma H_0 + \omega_{\boldsymbol{q}}$ and $\alpha$ being respectively the magnon frequency and Gilbert damping constant, and $D^R_{\boldsymbol{K}}(\nu) = (4\pi\hbar c^2/2\nu_{\boldsymbol{K}})[(\nu - \nu_{\boldsymbol{K}} + i0^+)^{-1} - (\nu + \nu_{\boldsymbol{K}} + i0^+)^{-1}]$ is the retarded component of the photon propagator. Integrating over $\omega$ by picking up the magnon poles and using the fact that the dominant contribution comes from a region $q \ll K_0$, $B_{\boldsymbol{k},\boldsymbol{q}}(\nu_{\boldsymbol{K}_0})$ is calculated to be

$$B_{\boldsymbol{k},\boldsymbol{q}}(\nu_{\boldsymbol{K}_0}) \approx -\frac{2\pi^2 \hbar \nu_{\boldsymbol{K}_0}}{cq\alpha} \delta\left(\hat{\boldsymbol{K}}_0 \cdot \hat{\boldsymbol{q}} - \frac{\omega_{\boldsymbol{q}}}{\nu_{\boldsymbol{q}}}\right) \left(\frac{1}{\omega_{\boldsymbol{q}}} \text{Im}\chi^R_{\boldsymbol{k}}(\omega_{\boldsymbol{q}})\right), \tag{S12}$$

where $\hat{\boldsymbol{K}}_0 = \boldsymbol{K}_0/K_0$ and $\hat{\boldsymbol{q}} = \boldsymbol{q}/q$. Therefore, the spin current pumped by light illumination is finally given by

$$J_S = \Gamma^+ \Gamma^- G_S \frac{\hbar \nu_{\boldsymbol{K}_0} a_S^4}{c\alpha} |\mathcal{E}_{\boldsymbol{K}_0}|^2, \tag{S13}$$

where $G_S = -\frac{\sqrt{2} N_{\text{int}} J_{\text{sd}}^2 S_0^2 \chi_P \tau_{\text{sf}}}{16\pi^2 \hbar^3 (\lambda_{\text{sf}}/a)^3} \Upsilon$ is a factor measuring the strength of the magnetic coupling at the PM/FI interface (corresponding to the spin mixing conductance), $a$ is the lattice constant of PM, and $\Upsilon = \int_0^1 dx \int_0^{a_S K_0} dy \frac{x^2}{(1+x^2)^2 + (2S_0 J_{\text{ex}}\tau_{\text{sf}}/\hbar)^2 y^2} \approx 0.142 a_S K_0$ with the strength of the exchange coupling $J_{\text{ex}}$ in FI and the dimensionless variables $x = \lambda_{\text{sf}} k$ and $y = \hbar\omega_{\boldsymbol{q}}/(2J_{\text{ex}}S_0)$. In equation (2) in the main text, for simplicity, we describe $\nu_{\boldsymbol{K}_0}$ and $\mathcal{E}_{\boldsymbol{K}_0}$ as $\nu$ and $E$, respectively. Equation (S13) represents a physical process in which magnons in FI is excited by a small energy transfer from external photons, thereby having the same sign as that for the ASP[41,42,*1]. This situation is consistent with the experimental results of the ASP in a Pt/$Y_3Fe_5O_{12}$ structure (e.g., Fig. 4d of ref. 41), where the sign of the background spin Seebeck signal owing to the heating of $Y_3Fe_5O_{12}$ is opposite to that of the ASP signal owing to the vibration of $Y_3Fe_5O_{12}$. Since the direction of the temperature gradient across the Pt/$BiY_2Fe_5O_{12}$ interface in the present experimental configuration is opposite to that in the ASP configuration, the voltage signal coming from the plasmonic spin pumping has the same sign as the background spin Seebeck signal (Fig. 2b).

As an endnote to this section, we mention the effect of the anisotropy of magneto-optical coefficients

---

[*1] Note that the definition of the voltage $V$ in this paper is opposite in sign to that in our previous literature on the ASP.

in $BiY_2Fe_5O_{12}$. According to Table 2 in Ref. 38, the magnitude of the anisotropy in magnetic linear birefringence is estimated to be at most 50 % in $Y_3Fe_5O_{12}$, which could in principle modulate the plasmonic-spin-pumping signal as the pumped spin current is proportional to $\Gamma^+\Gamma^-$ (see equation (S13)). However, such anisotropy effects are irrelevant to the observed voltage signals for the following reasons. Firstly, the anisotropy effects do not explain the experimental fact that the observed voltage signal is an odd function of the external magnetic field (Fig. 2d); the possible modulation of the plasmonic spin pumping induced by the anisotropy of magneto-optical coefficients must be an even function of the magnetic field. Secondly, in the present system, the polarization of near-field photons is distributed randomly, and thus such anisotropy effects are smeared out in the net signal. These arguments justify our conclusion that the voltage signal in the $Pt/BiY_2Fe_5O_{12}/Au$ NP sample is not affected by the anisotropy of magneto optical coefficients in $BiY_2Fe_5O_{12}$.

## Additional References